\documentclass[english,preprint,aps,a4paper]{revtex4-1}
\usepackage{ifpdf}
\usepackage[colorlinks,bookmarksopen]{hyperref}
\usepackage{babel}
\usepackage{ucs}
\usepackage[utf8x]{inputenc}
\usepackage[T1]{fontenc}
\usepackage{latexsym} 
\usepackage{graphicx} 
\usepackage{amsmath}
\usepackage{float}

\newcommand{\mean}[1]{
\left \langle #1 \right \rangle
}

\begin{document}

\title{Composition analysis based on Bayesian methods}
\author{G. Torralba Elipe and R.A. Vazquez }
\affiliation{Departamento de F\'{i}sica de
  Part\'{i}culas and\\ 
  Instituto Galego de F\'{\i}sica de Altas Enerx\'{\i}as, \\ 
  Campus Sur\\ 
  15782 Santiago de Compostela, Spain }
\date{\today}

\begin{abstract}
In this work we test the most widely used methods for fitting the composition
fraction in data, namely maximum likelihood, $\chi^2$, mean value of the
distributions and mean value of the posterior probability function.  We
discuss the discrimination power of the four methods in different scenarios:
signal to noise discrimination; two signals; and distributions of Xmax for
mixed primary mass composition. We introduce a "distance" parameter, which can
be used to estimate, as a rule of thumb, the precision of the discrimination.
Finally, we conclude that the most reliable methods in all the studied
scenarios are the maximum likelihood and the mean value of the posterior
probability function.
\end{abstract}

\maketitle
%\linenumbers

\section{Introduction}
One of the main challenges of the astroparticle physics reasearch is to
determine the chemical composition of the cosmic rays that reach the earth.
At the highest cosmic ray energies, analysis of composition must be made with
indirect methods by using the evolution of cosmic ray showers in the
atmosphere.  In this note, we use four different methods to estimate the
composition given a single variable (e.g. the $X_{\rm max}$ distribution) by
using Bayesian methods. A similar work, using Monte Carlo techniques, to
study the efficiency of different discriminators can be found in \cite{Durso}.

The note is as follows: in section \ref{Methods} we discuss the four different
methods and we apply them in simple analytical cases in section
\ref{Analytical}. In section \ref{Examples}, we apply the different methods to
two physical cases Signal to Noise discrimination and two overlapped
signals. Finally, we evaluate the methods for the specific example of $X_{\rm
  max}$ distributions with realistic probability densities and we get a
possible composition of the cosmic rays in section \ref{Xmax}.

To be concrete, we will consider a two composition scenario. Although the
methods discussed can be easily generalized to include more than two
distributions. 

\section{Methods}
\label{Methods}

Consider the following problem. A given data variable is extracted from two
different probability distributions with a ``composition'' fraction $\alpha$,
($0 \le \alpha \le 1$) so that the joint probability distribution is given by
\begin{equation}
f(x;\alpha) = \alpha g_1(x) + (1-\alpha ) g_2(x).
\label{data_dist}
\end{equation}
The probability distributions $g_{1,2}(x)$ are known and the problem consists
in  determining the composition fraction $\alpha$ from the measurement of 
$n$ data points $x_i$, $i= 1,\cdots ,n$.
If $\alpha$ was known, the probability of getting the data $D=\{x_i\}$ is
given by
\begin{equation}
P(D|\alpha I) = \prod_{i=1}^n f(x_i;\alpha), 
\label{p_data}
\end{equation}
where $I$ is any prior information we have about the problem, including the
prescription of the probabilities $g_i$. Here we are implicitly assuming that
the different data points are independent.  Using Bayes's theorem \cite{Evans},
we can obtain the posterior probability for $\alpha$ given the data
\begin{equation}
P(\alpha | D I) = \frac{P(D |\alpha I) P(\alpha | I)}{P(D|I)}.
\label{bayes}
\end{equation}
$P(D|I)$ is the probability of obtaining the given data independently of any
value of $\alpha$ and here acts as a normalization constant. $P(\alpha|I)$ is
the prior probability for $\alpha$. In our problem, astrophysical input may
give information on the cosmic ray composition and give preference for, say,
proton domination. In the absence of any information a flat distribution gives
good results. In the following we will use $P(\alpha|I) = 1$ but all our
results will be valid for other choices of prior probabilities.  
Therefore, we can write  Eq.(\ref{bayes}) as
\begin{equation}
P(\alpha | D I) = \frac{1}{\mathcal{N}}\prod_{i=1}^{N} \left \{\alpha g_1(x_i) +
  (1-\alpha ) g_2(x_i)\right \},
\label{prob}
\end{equation}
where $\mathcal{N}$ is a normalization constant. In some problems instead of
Eq. (\ref{prob}), where all the data points are given, one has data binned in
the variable $x$. In that case the equation reads
\begin{equation}
P(\alpha | D_k I) = \frac{1}{\mathcal{N}}\prod_{j=1}^{k}  \left \{\alpha
G_1(x_j) + (1-\alpha ) G_2(x_j) \right \}^{n_k},
\label{prob_bin}
\end{equation}
where the data now is $D_k = \{ n_1,\cdots,n_k\}$, the number of events in the
bins $1,\cdots,k$ with center values $x_1,\cdots,x_k$, and $G_i(x_j) =
\int_{x_j} g_i(x) dx $ is the integral on the bin of the probability
density. Although binning the data makes the problem somehow easier, it wastes
information. 
 
Eq. (\ref{prob}) (or alternatively Eq. \eqref{prob_bin}) contains all the
information we have about our problem.  Estimation of the composition fraction
reduces to the choice of a ``best estimator''. Now we examine the
different choices which are currently used in the literature.

\begin{itemize}

\item $\alpha_{<>}$

We can calculate the mean of the data points and choose $\alpha$ such that
it coincides with the mean value of the distributions, {\it i.e.}
if $\bar x = \frac{1}{n} \sum_{i=1}^n x_i$, then 
\begin{equation}
\bar x = \alpha \mean{g_1(x)} + (1-\alpha )\mean{g_2(x)}
\label{means}
\end{equation}
if we define $\bar x_i= \int x g_i(x) dx $, then we get
\begin{equation}
\alpha_{<>} = \frac{ \bar x- \bar x_2}{ \bar x_1 - \bar x_2}.
\label{alpha_means}
\end{equation}
Eq. (\ref{alpha_means}) has a simple analytical form and is easy to evaluate
for any distribution (provided it has first moments). This is a useful
advantage when working with the first few moments (see e.g.
\cite{Abreu}). However, it can give unphysical results (($\alpha >1$,
$\alpha<0$) and gives the largest deviation with respect to the true value for
all studied estimators.

\item $\alpha_{\rm max}$

Alternatively, one can choose as the best estimator, the value that maximizes
the posterior probability.  
\begin{equation}
\frac{\partial P(\alpha |D I)}{\partial \alpha} = 0.
\label{ml}
\end{equation} 
Since $P(\alpha)$ is positive an equivalent alternative is to maximize the
logarithm of $P$. Then
\begin{equation}
\frac{\partial \log P(\alpha |D I)}{\partial \alpha} = \sum_{i=1}^n
\frac{g_1(x_i)-g_2(x_i)}{\alpha g_1(x_i) + (1-\alpha ) g_2(x_i)} =0.
\label{mllog}
\end{equation} 
This is the maximum likelihood estimation. It is known to give very good
estimation of $\alpha$ in almost all cases, even for small number of events.
It has the disadvantage that an analytical solution is possible only for very
small number of events or bins. This method is used, for instance, in the
standard package TFractionFitter of ROOT \cite{TFitter}. Usually, the solution is
found by numerically searching for the maximum of Eqs. (\ref{ml},\ref{mllog}).

\item $\alpha_{\chi}$

If the number of events is large, one expects a well defined peak
distribution in $\alpha$. Near the maximum of the distribution one can
approximate this distribution by a gaussian.  Binning the data
we can construct a $\chi^2$ variable for
the problem
\begin{equation}
\chi^2(\alpha ) = \sum_{j=1}^{k} \frac{(n_j/n - F(x_j,\alpha))^2}{n_j},
\label{chi}
\end{equation}
where $n_j$ is the number of data events in bin $j$ and $F(x_j,\alpha)= \alpha
G_1(x_j) + (1-\alpha) G_2(x_j)$
is the probability of having an event in bin $j$ for a given $\alpha$. 
The optimal value of $\alpha$ can be found minimizing the $\chi^2$
\begin{equation}
\frac{\partial \chi^2 (\alpha)}{\partial \alpha} = 0.
\label{chimin}
\end{equation}
The solution of Eq. (\eqref{chimin}) has the advantage of being an analytical
and relatively simple expression
\begin{equation}
\alpha_{\chi} = \frac{\sum_{j=1}^{k} (n_j/n -G_2(x_j))(G_1(x_j)-G_2(x_j))}
      {\sum_{j=1}^{k}(G_1(x_j)-G_2(x_j))^2}.
\label{alphachimin}
\end{equation}
It is an asymptotic limit (for $n$ and $n_j$ large) of the maximum likelihood
method and, as such, gives very good results in this limit. 

\item $\left \langle \alpha \right \rangle$

Once we know the probability density function for $\alpha$, we can obtain the
mean value of the distribution
\begin{equation}
\left \langle \alpha \right \rangle = \int_0^1 d\alpha \; \alpha \; 
P(\alpha|D I).
\label{means_cont}
\end{equation}
Although this estimator is not much used, we will shown below that it gives
the best performance in most cases. It has the disadvantage of being
difficult to evaluate analytically but for the simplest cases.

\item $\alpha_M$

Any other estimator could provide sensible results. As an example, the median
of the posterior probability, defined by 
\begin{equation}
\int_0^{\alpha_M}  d\alpha  P(\alpha |D I) = \int_{\alpha_M}^1  d\alpha
P(\alpha |D I) = 1/2. 
\label{alphamedian}
\end{equation}
It is well known that the median is a robust estimator, being invariant
against a large set of transformations of the probability distributions. 
However, it is difficult to evaluate both analytically and
numerically. Therefore we do not consider it any further. 

\end{itemize}

\subsection{A toy analytical case: heads or tails}
\label{Analytical}

The simplest problem of discrimination is the following: Assume that the two
distributions $g_1(x)$ and $g_2(x)$ are totally separated ({\it i.e.} they 
do not overlap). In that case, the actual shape of $g_1$ and $g_2$ is
irrelevant and one can bin the data in just two bins $x=a,b$ such that all the
probability is concentrated in either bin $a$ or $b$. So, let the two
probability functions be
\begin{equation}
G_1(x) = \left \{ \begin{matrix} 1 & \mbox{if } x = a
\\ 0 & \mbox{if } x = b\end{matrix}\right.  
\label{dist11}
\end{equation}
\begin{equation}
G_2(x) = \left \{ \begin{matrix} 0 & \mbox{if } x = a
\\ 1 & \mbox{if } x = b\end{matrix}\right.  
\label{dist12}
\end{equation}

Denoting $\phi(\alpha) = P(\alpha | D I)$ and applying Eq. (\ref{prob}), then
\begin{equation}
\phi(\alpha) = \frac{1}{\mathcal{N}} \left [ \alpha^n(1-\alpha)^{N-n} \right ]
\label{prob_ana}
\end{equation}
where $n$ is the number of events with $x=a$ and $N-n$ is the number of
events with $x=b$, and $N$ the total number of events. This is just a problem
of determining the probability of having heads or tails in a (possibly) loaded
coin given the number of heads and tails in an experiment. As could be
expected, it is just a binomial distribution. 

By direct calculation, one obtains
\begin{equation}
\alpha_{\chi} = \alpha_{\rm max} = \alpha_{<>} = \frac{n}{N}.
\label{coinbest}
\end{equation}
We obtain as the estimation of the probability of ``head''
events, the fraction of head events observed. On the other hand the mean
$\alpha$ gives 
\begin{equation}
\left \langle \alpha \right \rangle =  \frac{n+1}{N+2}. 
\label{coinLaplace}
\end{equation}
Although this result may be surprising at first sight it is a well known
result in the literature. It is known as Laplace's succession rule. One may
notice that in the limit $N,n \rightarrow \infty$ with $n/N$ fixed one
recovers Eq. \eqref{coinbest}. Note that if $N=0$, then $n=0$ and all the
methods are indefinite except the mean value which gives $\frac{1}{2}$.  This
is just the mean value of the prior probability. If $N=1$, then either $n=0$
or $n=1$, which would give either $\alpha_{\rm max} = 0$ or
$1$. Eq. \eqref{coinLaplace} gives $\left \langle \alpha \right \rangle = 1/3$
or $2/3$.  In section \ref{Examples} we show numerically this phenomenon for a
more realistic model.

\subsection{Heads or tails with contamination}
\label{anaII}

For a more interesting case, consider now the previous example but with a
(possibly small) contamination between both distributions
\begin{equation}
G_1(x) = \left \{ \begin{matrix} 1-\epsilon & \mbox{if } x = a
\\ \epsilon & \mbox{if } x = b\end{matrix}\right.  
\label{cont1}
\end{equation}
\begin{equation}
G_2(x) = \left \{ \begin{matrix} \delta  & \mbox{if } x = a
\\ 1-\delta & \mbox{if } x = b\end{matrix}\right.  
\label{cont2}
\end{equation}
So that there is a (small) probability of a event of type 1 (``heads'') to be
identified in the bin 2 (``tails'') and vice-versa.  The posterior probability,
after measuring $N= n_1+n_2$ total events with $n_1$ of type 1 and $n_2$ of
type 2 is
\begin{equation}
\phi(\alpha) = \frac{1}{\mathcal{N}} \left [ \alpha (1-\epsilon) + (1-\alpha)
  \delta  \right ]^{n_1}  
\left [ \alpha \epsilon + (1-\alpha)
  (1-\delta)  \right ]^{n_2}. 
\label{prob_ana_cont}
\end{equation}
After some algebra one obtains again
\begin{equation}
\alpha_{\chi} = \alpha_{\rm max} = \alpha_{<>} =
\frac{1}{1-\delta-\epsilon} \left [ \frac{n_1}{N} - \delta \right ].
\label{coinbest_cont}
\end{equation}
The mean value of $\alpha$ has not a simple analytical expression. It is given
by
\begin{equation}
\left \langle \alpha \right \rangle = \frac{1}{1-\delta-\epsilon} \left(
\frac{B(1-\epsilon,n_1+2,n_2+1)-B(\delta,n_1+2,n_2+1)}
{B(1-\epsilon,n_1+1,n_2+1)-B(\delta,n_1+1,n_2+1)} \right)
-\frac{\delta}{1-\delta-\epsilon},
\end{equation}
where $B(x,n_1,n_2)$ is the incomplete Beta function \cite{Abramowitz}
\begin{equation}
B(x,n_1,n_2) = \int_0^x dy y^{n_1-1} (1-y)^{n_2-1}.
\end{equation}
For $n_1$ and $n_2$ integers $B$ is a polynomial in $x$. One can show that the
above equation gives always physical values $0 \le \alpha \le 1$, even for
degenerate cases. 

Although this model is rather simplistic, it has all the ingredients found in
actual cases. One can interpret Eq. \eqref{coinbest_cont} rather easily, the
term $-\delta$ subtract the expected fraction of events of type 2 which
fall into bin 1. On the other hand the factor $1-\delta-\epsilon$ is a measure
of the fraction of well identified events. It is also a measure of the
overlapping of the two distribution. As we will see below, this is a general
characteristic of the problem. 
Another interesting point of Eq. \eqref{coinbest_cont} is the fact that it can
produce unphysical results. If $n/N < \delta$, the expected fraction is
negative. This is so because even for $\alpha=0$, we expect a number of events 
in the first bin of $\delta N$. 
Finally, one can see that the case $\epsilon+\delta=1$ is ill defined. But in
this case both distributions are equal: no discrimination
can be made between the two distributions. 

The mean value determination $\left \langle \alpha \right \rangle$ does not
suffer from this behavior, always giving physically admissible results. In the
case of the two distributions being equal, we would obtain $\left \langle
\alpha \right \rangle = 1/2$, which is easily interpreted. If the data can not
differentiate between the two cases we do not gain any information from the
data and the estimation given by our prior is kept.

\section{Application of the methods} \label{Examples}

We now apply the methods discussed previously to several different scenarios.
In section \ref{signal_noise}, we study a typical problem of signal/noise 
identification. In section \ref{two_signals}, we concentrate on the separation
of two signals and we study the dependence of the resolution with respect to
the distance between the two signals. 

A number of distance measures for probabilities has been proposed in the
literature. In the appendix \ref{appendixA} we discuss some possibilities and
justify the choice of the overlapping area, as our distance.
Given two distributions $g_1(x)$ and $g_2(x)$ we define the distance between
the two distribution as
\begin{equation}
d_1(g_1,g_2) = \int dx |g_1(x)-g_2(x)|.
\end{equation}
Which ranges between 0 and 2. For $d_1=2$ the distributions do not overlap;
for $d_1=0$ the distributions are equal. For the previous example of heads and
tails, the distance is given by $d_1 = 2(1-\delta-\epsilon)$, which is the
pre-factor appearing in \eqref{coinbest_cont}.

\subsection{Signal/Noise discrimination}
\label{signal_noise}

Consider the case of extracting a signal with a well defined peak from events
coming from the signal plus a flat noise. To be concrete, we will choose the
following probability density functions
\begin{equation}
g_1(x) = \frac{1}{\cal N} \exp{  \left \{-\frac{(x-\mu)^2}{2\sigma^2}  
\right \}} ; x \in [a+\epsilon,b-\epsilon],
\end{equation}
\begin{equation}
g_2(x) = \frac{1}{b-a};   x \in [a,b].
\end{equation}
Here $[a,b]$ is the range of the variable. We take the signal to be different
from zero in a subrange of this interval, defined by $\epsilon>0$. $\cal N$ is
a normalization constant and $\mu$ and $\sigma$ are the mean and RMS of the
gaussian. In the numerical calculations we will choose $a=0$, $b=7$, $\epsilon
= 1$, $\mu = 2$, and $\sigma = 0.2$. The distance $d_1$ for this case is
$1.694$.

\begin{center}
\begin{figure}[H]
\centering
\includegraphics[scale=0.5]{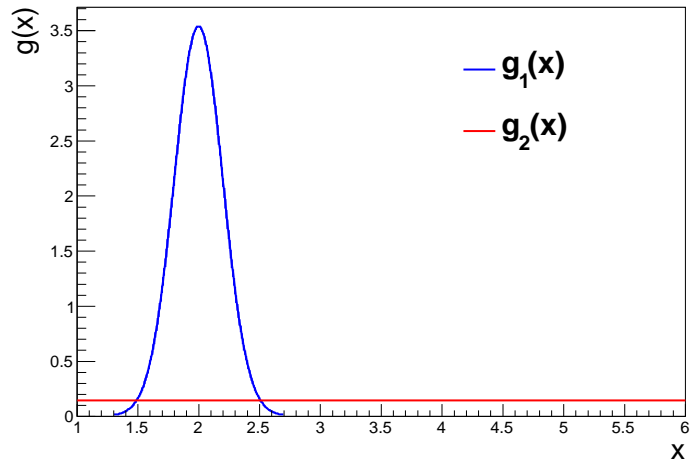}
\caption{Probability density functions for signal and noise.}
\label{snpdfs}
\end{figure}
\end{center}

\begin{center}
\begin{figure}[H]
\centering
\includegraphics[width=0.7\textwidth]{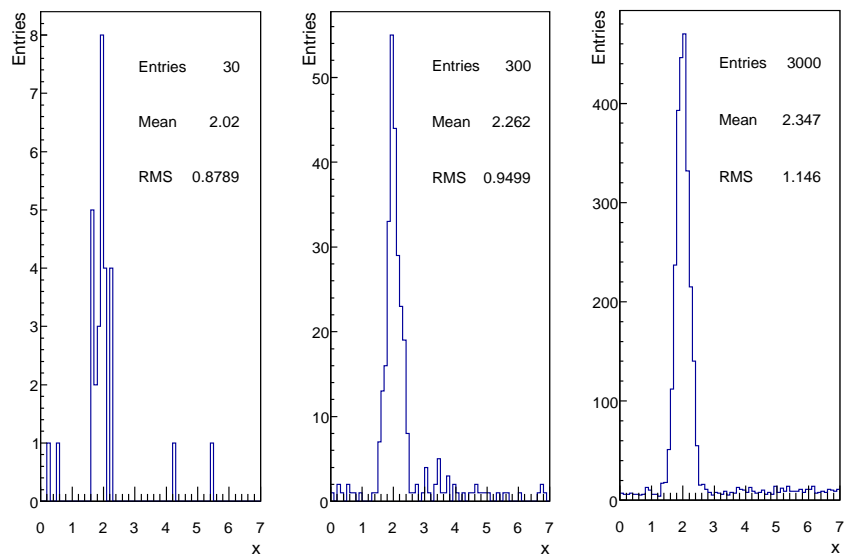}
\caption{Data histograms for 30, 300 and 3000 events sampled from the
  distributions in Fig.\ref{snpdfs}.}
\label{sndatas}
\end{figure}
\end{center}

\begin{figure}[H]
\centering
\includegraphics[width=1\textwidth]{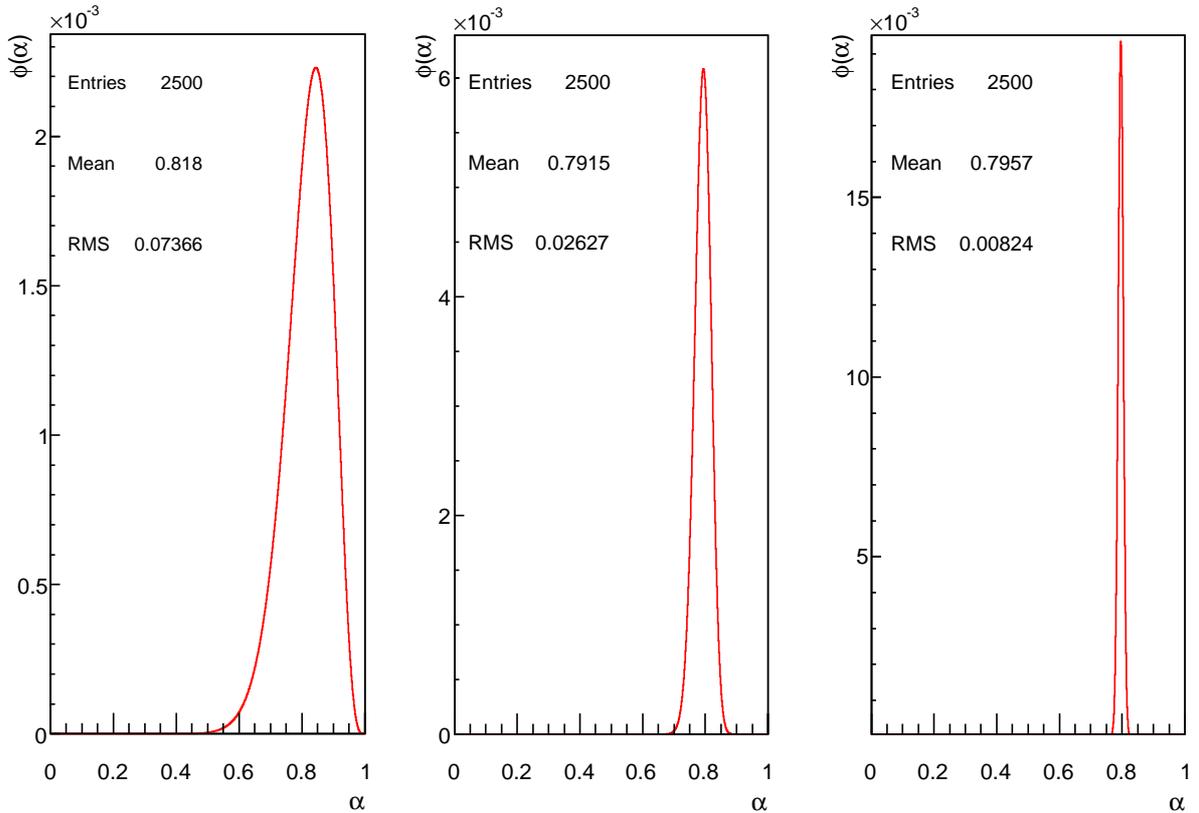}
\caption{Probability functions of $\alpha$ for 30, 300 and 3000 events. The
  true fraction is 0.8}
\label{snfs}
\end{figure}

\begin{figure}[H]
\includegraphics[width=1\textwidth]{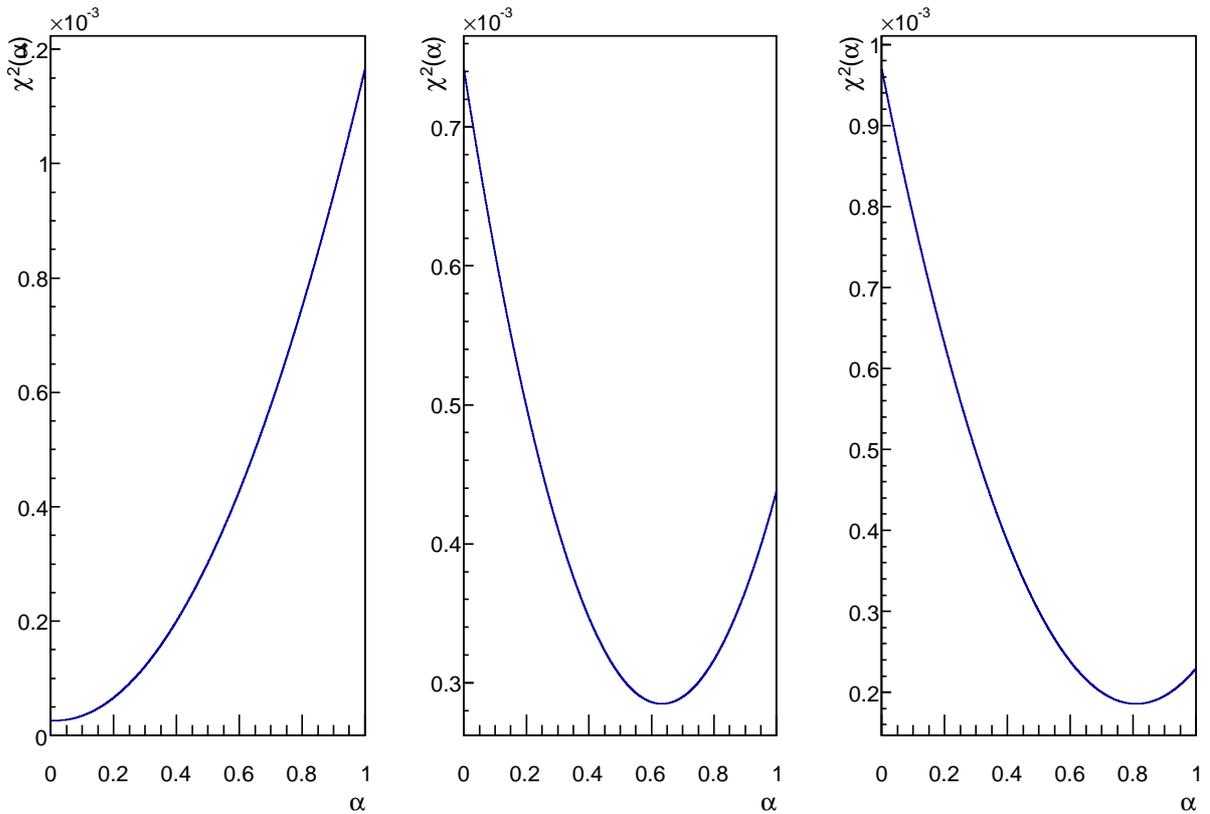}
\caption{$\chi^2 $ distributions for fixed number of events of 30, 300, and
  3000. The true fraction is 0.8.}
\label{snchis}
\end{figure}

In Fig. (\ref{snpdfs}) we show both probability density functions.
As a first numerical evaluation we calculate the estimated fraction for a true 
signal fraction of $\alpha = 0.8$ with a fixed number of events of 
30, 300 and 3000. In Fig. \ref{sndatas} we show the data in a typical
run. 
In table \ref{table:signal-noise} we show the results for all estimators
discussed and for the cases with different number of events. In Figures
\ref{snfs} and \ref{snchis} we show the fraction probability and $\chi^{2}$
functions obtained.
\begin{table}[h]
\begin{center}
\begin{tabular}{ccccc}
\hline
\hline
\# Events  & $\mean{\alpha }$ & $\alpha_{\rm max}$ 
& $\alpha_{\chi^2}$ & $\alpha_{<>}$   \\
\hline
\hline
30 & 0.82 & 0.84 & 0.00 & 0.70 \\
\hline
300 & 0.79 & 0.79 & 0.63 & 0.82 \\
\hline
3000 & 0.80 & 0.80 & 0.81 & 0.78 \\
\hline
\hline
\end{tabular}
\caption{Results for the signal/noise discrimination.}
\label{table:signal-noise}
\end{center}
\end{table}
Note that all of the methods give a reasonable fraction, but the mean value
gives the estimated fraction closest to the true fraction. In this case, we
can not choose a method or another, getting the same results except the
$\chi^2$ method, which is the worst estimator for this example. 
For the uncertainties we may take the RMS of the posterior probability
functions. We obtain 0.07, 0.03 and 0.008 for 30, 300 and 3000 events. 
The $\chi^2$ estimator gives such a bad result due to the chosen binning. We
have chosen a bin size of 0.002, which for small number of events is
unreasonable. This was done on purpose to show that one does not need to bin
the data and that binning can produce bad results, if poorly done. 

\subsection{Two signals}
\label{two_signals}

As a further example consider now the problem of discrimination of two signals
which we will model as gaussians
\begin{equation}
g_1(x) = \frac{1}{\sigma_1 \sqrt{2\pi}} \exp{  \left \{-\frac{(x-\mu_1
    )^2}{2\sigma_1^2}  \right \}},
\end{equation}
\begin{equation}
g_2(x) = \frac{1}{\sigma_2 \sqrt{2\pi}} \exp{  \left \{-\frac{(x-\mu_2
    )^2}{2\sigma_2^2}  \right \}}.
\end{equation}
We will use $\mu_1 = 0$, $\sigma_1 = 1$, and $\sigma_2 = 0.5$. In our
numerical examples, we vary $\mu_2$ from -1.5 to 1.5 to search for the
efficiency of the methods for different distances between the probability
distributions and we will also use different values of the composition
fraction.

For gaussian distributions the distance between the two functions can be
written as
\begin{equation}
d_1(g_1,g_2) = \int dx |g_1(x) -g_2(x)| = I_1 + I_2 + I_3,
\end{equation}
where the $I_i$ are combinations of error functions
\begin{equation}
I_1 = \frac{1}{2} |{\rm Erf}(\frac{x_{c1}-\mu_1}{\sqrt(2) \sigma_1})-{\rm
  Erf}(\frac{x_{c1}-\mu_2}{\sqrt(2) \sigma_2}) |,
\end{equation}
\begin{equation}
I_2 = \frac{1}{2} |{\rm Erf}(\frac{x_{c2}-\mu_1}{\sqrt(2) \sigma_1})-{\rm
  Erf}(\frac{x_{c1}-\mu_1}{\sqrt(2) \sigma_1}) -
{\rm Erf}(\frac{x_{c2}-\mu_2}{\sqrt(2) \sigma_2})+{\rm
  Erf}(\frac{x_{c1}-\mu_2}{\sqrt(2) \sigma_2})|,
\end{equation}
\begin{equation}
I_3 = \frac{1}{2} |{\rm Erf}(\frac{\mu_1-x_{c2}}{\sqrt(2) \sigma_1})-{\rm
  Erf}(\frac{\mu_2-x_{c2}}{\sqrt(2) \sigma_2}) |,
\end{equation}
and where $x_{c1,2}$ are the two solutions to the equation
$$
g_1(x) = g_2(x).
$$

\begin{figure}[H]
\centering
\includegraphics[width=0.7\textwidth]{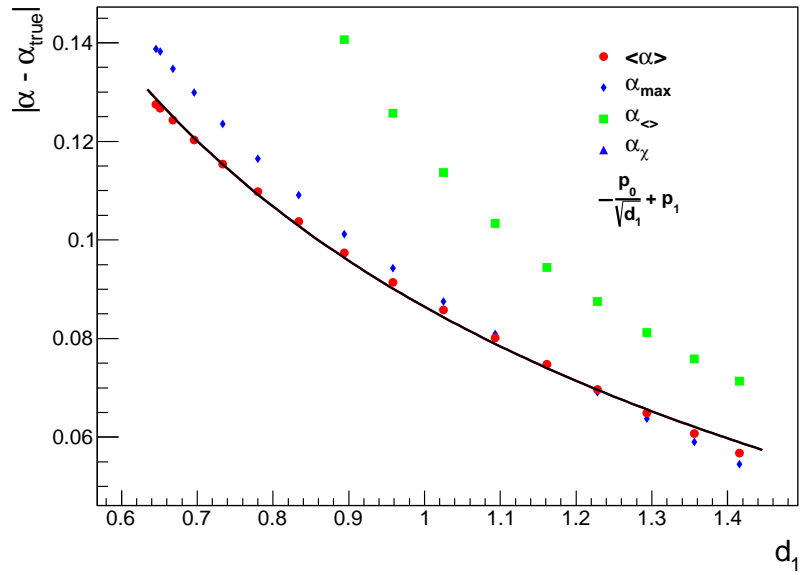}
\caption{Absolute difference between the estimated fraction and true fraction
  for 30 data sample as a function of the distance. Note that $d_1=0$ means
  that the two distributions are equal while $d_1=2$ means that the
  distributions are completely separated.}
\label{dist30}
\end{figure}

\begin{figure}[H]
\centering
\includegraphics[width=0.7\textwidth]{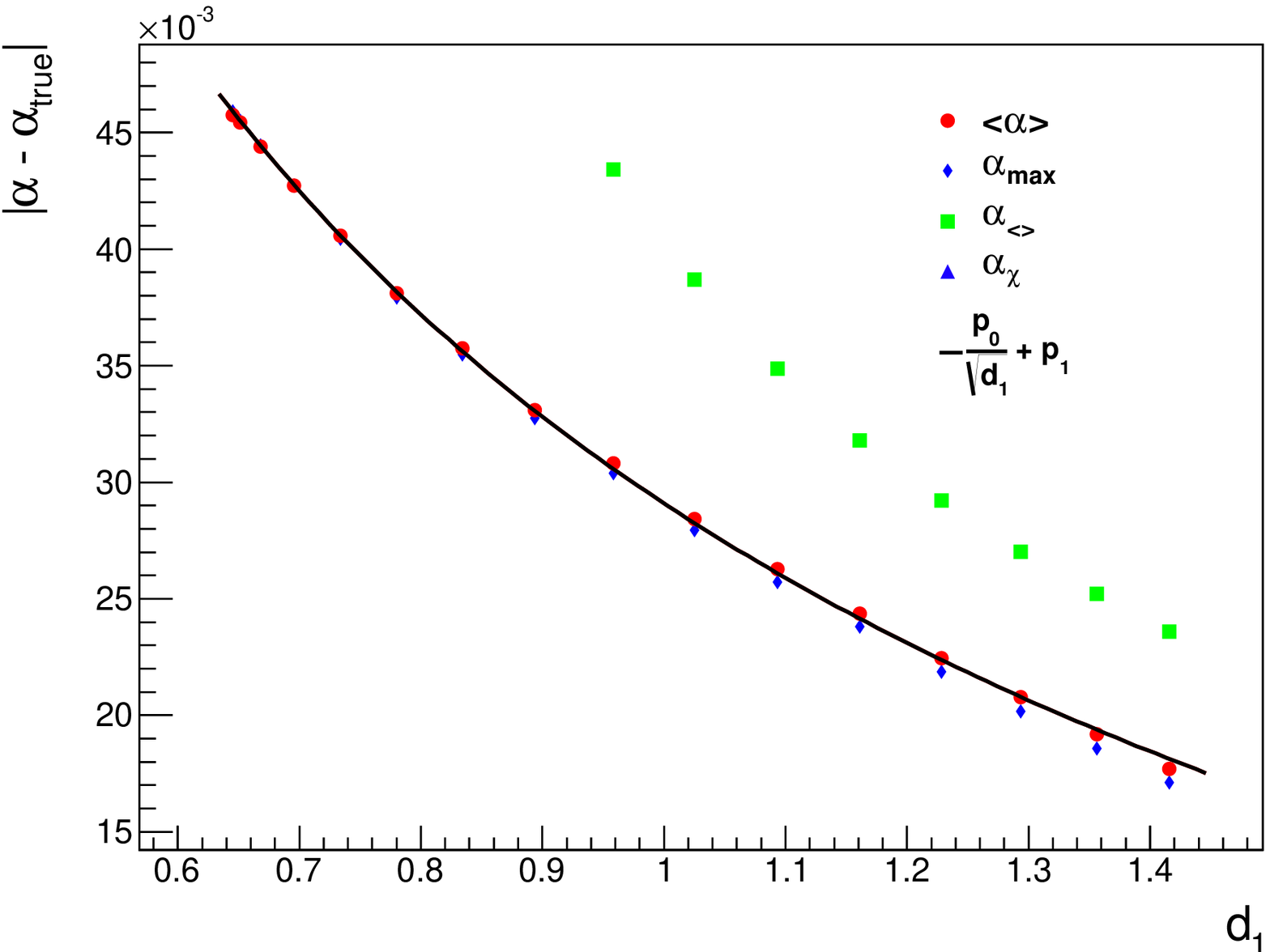}
\caption{Absolute difference between the calculated fraction and true fraction
  for 300 data sample as a function of the distance.}
\label{dist300}
\end{figure}

\begin{figure}[H]
\centering
\includegraphics[width=0.7\textwidth]{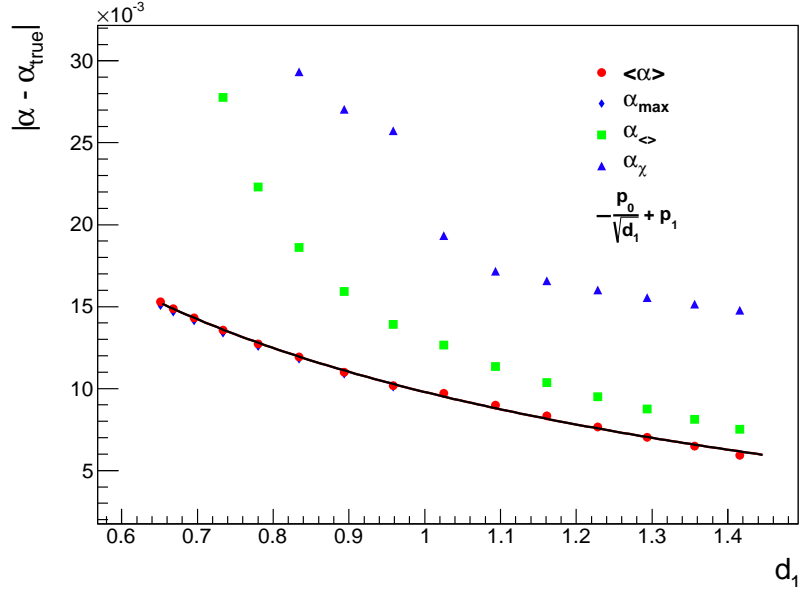}
\caption{Absolute difference between the calculated fraction and true fraction
  for 3000 data sample as a function of the distance.}
\label{dist3000}
\end{figure}

We have run 10000 trials with 30, 300, and 3000 events for each value of the
composition fraction, $\alpha_{true}$, varying from 0 to 1 in steps of 0.1.
In figures \ref{dist30}-\ref{dist3000} we show $|\alpha - \alpha_{true}|$
as a function of the distance for 30, 300, and 3000 events. Note that the best
estimator is the mean value of the posterior probability. For a large number
of events this estimator tends to the maximum likelihood, but for small number
of events or small distances it performs slightly better. 

\begin{figure}[H]
\centering
\includegraphics[width=0.85\textwidth]{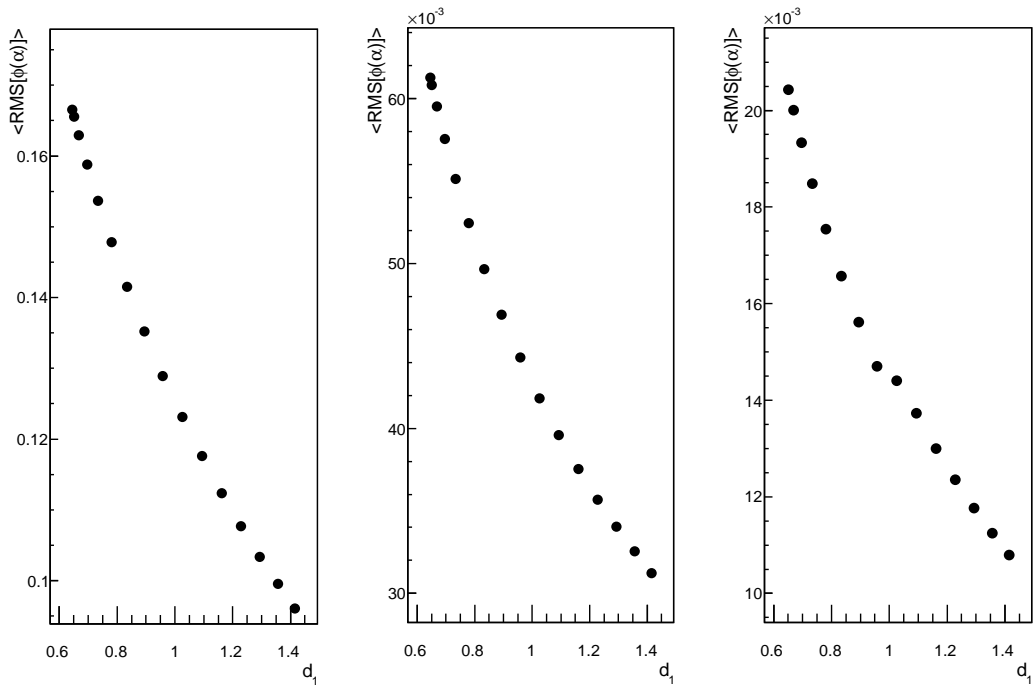}
\caption{RMS of the posterior probability distribution for all trials as a
  function of the distance between the probability distributions. Left:
  analysis done with 30 events. Middle: 300 events. Right: 3000 events.}
\label{sigmas}
\end{figure}

The $\chi^2$ method does not appear in figures \ref{dist30} and \ref{dist300},
it is off scale due to the binning used in the estimation of the $\chi^2$. In
figure \ref{dist3000} it is the worst method for the same reason.  
One can see in the same figures
figures \ref{dist30}-\ref{dist3000} that both $\mean{\alpha }$ and
$\alpha_{max}$ scale as the squate root square of the distance. A
fit to the function 
\begin{equation}
|\alpha -\alpha_{true}| = \frac{p_0}{\sqrt(d_1)}+p_1,
\end{equation}
is shown in the figures.  This is in agreement with the results of section
\ref{anaII} and confirms our choice for the distance.  The uncertainty of
$\alpha_{<>}$ more than double that of the maximum likelihood or the mean
value for small distances.  The corresponding RMS of the posterior probability
distribution are shown in figure \ref{sigmas}.

We now apply the methods for the gaussians with different mean and standard
deviation for a single trial. In this case, we choose $\mu_1 = 2$, $\sigma_1 =
0.2$, $\mu_2 = 2.3$ and $\sigma_2 = 0.4$.  The distance between the
distributions is $d_1 = 0.926$. By looking at figures
\ref{dist30}-\ref{dist3000}, we expect the fraction to be estimated with an
uncertainty of $\sim 0.1$, $0.032$ and $0.001$ for 30, 300 and 3000 events
respectively for the $\mean{\alpha}$ or $\alpha_{max}$ estimators. For the
$\alpha_{<>}$ estimator we expect the uncertainty to be $< 0.14$, $0.045$ and
$0.014$.

\begin{figure}[H]
\centering
\includegraphics[scale=0.5]{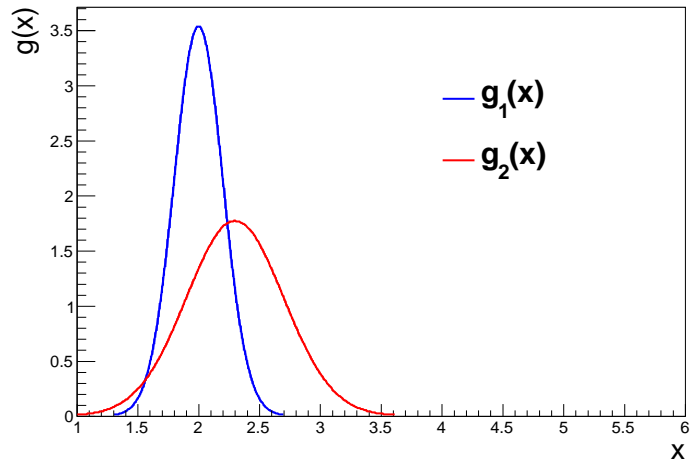}
\caption{Probability density functions}
\label{tspdfs}
\end{figure}

\begin{figure}[H]
\centering
\includegraphics[scale=0.6]{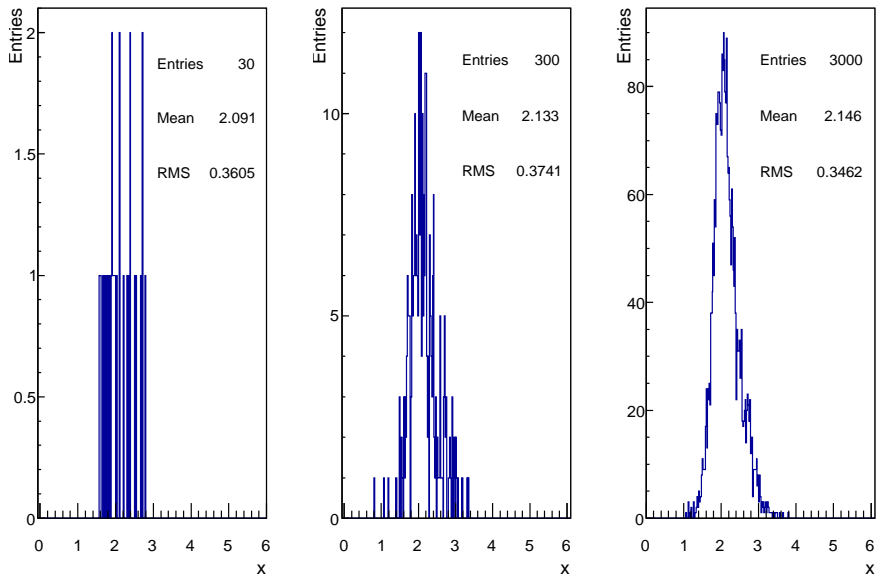}
\caption{Data distributions analyzed for 30, 300 and 3000 events corresponding
to the probability distributions in Fig. \ref{tspdfs}.}
\label{tsdatas}
\end{figure}

\begin{figure}[H]
\centering
\includegraphics[scale=0.6]{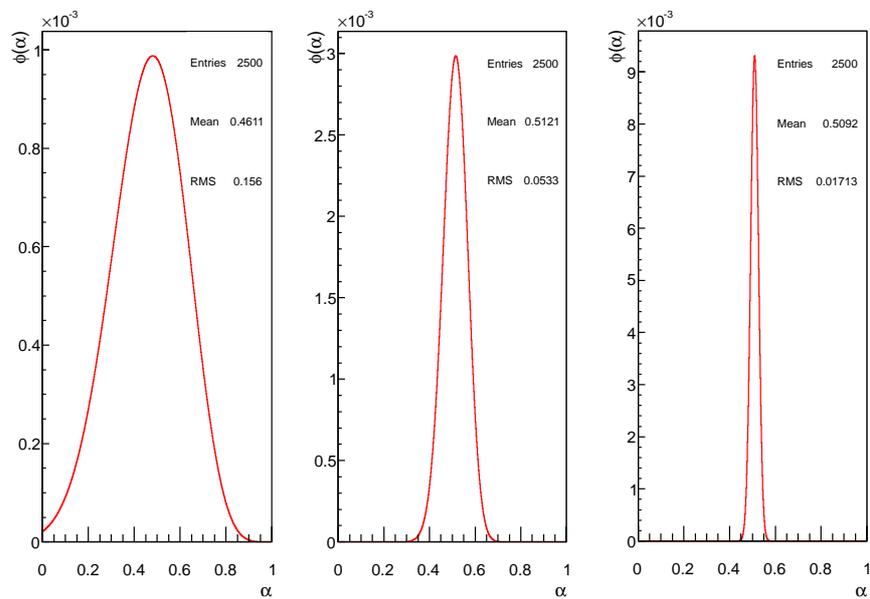}
\caption{Posterior probability distributions of $\alpha$ for 30, 300, and 3000
  events. The true fraction is $\alpha_{true} = 0.5$.}
\label{tsfs}
\end{figure}

\begin{figure}[H]
\centering
\includegraphics[scale=0.6]{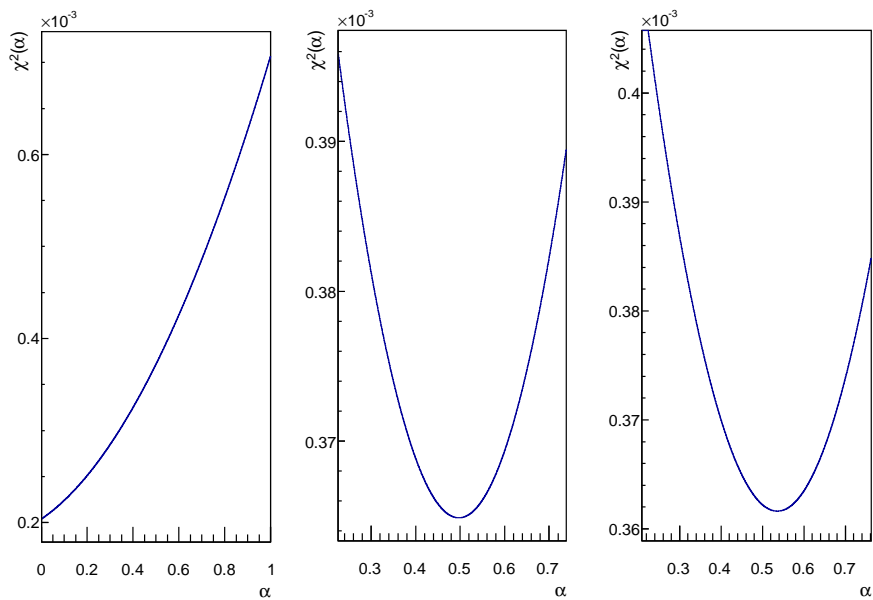}
\caption{$\chi² $ functions for the same cases as figure \ref{tsfs}.}
\label{tschis}
\end{figure}

In figure \ref{tspdfs} we show the probability density functions for
$g_1(x)$ and $g_2(x)$ considered here. Examples of data
distributions are shown in figure \ref{tsdatas}.
In table \ref{table:gauss} we show the results for the four methods and
figures \ref{tsfs} and \ref{tschis} we show the probability 
distributions and the $\chi^2$ distributions.
\begin{table}[H]
\centering
\begin{tabular}{ccccc}
\hline
\hline
\# Events & $\mean{\alpha }$ & $\alpha_{\rm max}$ & $\alpha_{\chi^2}$ &
$\alpha_{<>}$   \\ 
\hline
\hline
30 &  0.46 & 0.48 & 0.0 & 0.70 \\
\hline
300 & 0.51 & 0.52 & 0.50 & 0.55 \\
\hline
3000 & 0.51 & 0.51 & 0.54 & 0.51 \\
\hline
\hline
\end{tabular}
\label{table:gauss}
\caption{Results of the methods for the data samples. The true fraction is
  $\alpha_{true} = 0.5$.}
\end{table}

\section{Analysis of composition using $X_{\rm max}$ distributions} 
\label{Xmax}

In the previous sections we have shown that the best estimators for the
fraction are the mean value and maximum value of the probability distribution
$P(\alpha | D I)$. We will now make an analysis of composition of the high
energy cosmic rays using the maximum of the longitudinal profile, $X_{\rm
  max}$ as our discriminator \cite{Durso}.  We will use all the estimators
discussed previously but concentrate on the results of the mean value and the
maximum likelihood. The $X_{\rm max}$ distributions are generated with the
CONEX generator using EPOS as the hadronic model.

\begin{figure}[H]
\centering
\includegraphics[scale=0.5]{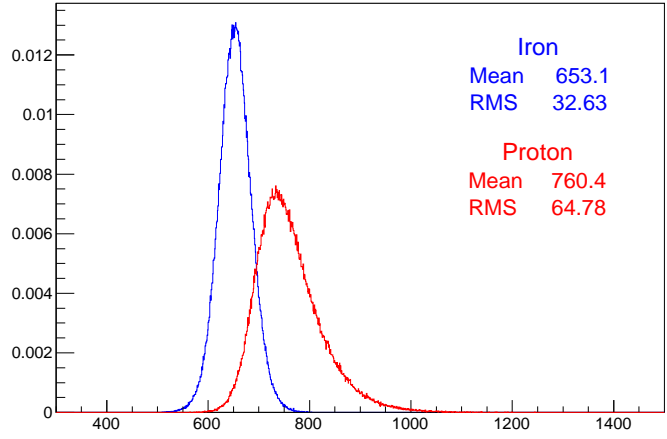}
\caption{$X_{\rm max}$ distributions for Iron and Proton obtained from
  CONEX-EPOS with a primary energy from 1  EeV to 3.16 EeV.}
\label{xmaxpdfs}
\end{figure}

Consider a typical example, where we want to find the proton and iron fraction
in a data sample corresponding to energies between 1 EeV to 3.16 EeV.  The
distributions of $X_{\rm max}$ are shown in figure \ref{xmaxpdfs}.  The
distance $d_{1}$ between the simulated distributions with the EPOS model in
this energy bin is $1.55$, then, we can use, as a rule of thumb, our estimated
resolution in the composition fraction to about $|\alpha-\alpha_{\rm true}|
\sim 1/\sqrt{N d_1}$ which amounts to 0.05, 0.012, or 0.006 for 30, 300, or
3000 events (see figures \ref{dist30}-\ref{dist3000}). The expected RMS of the
distributions will be of order $\sigma \sim 0.1, 0.03, 0.01$ respectively.

\begin{figure}[H]
\centering
\includegraphics[scale=0.6]{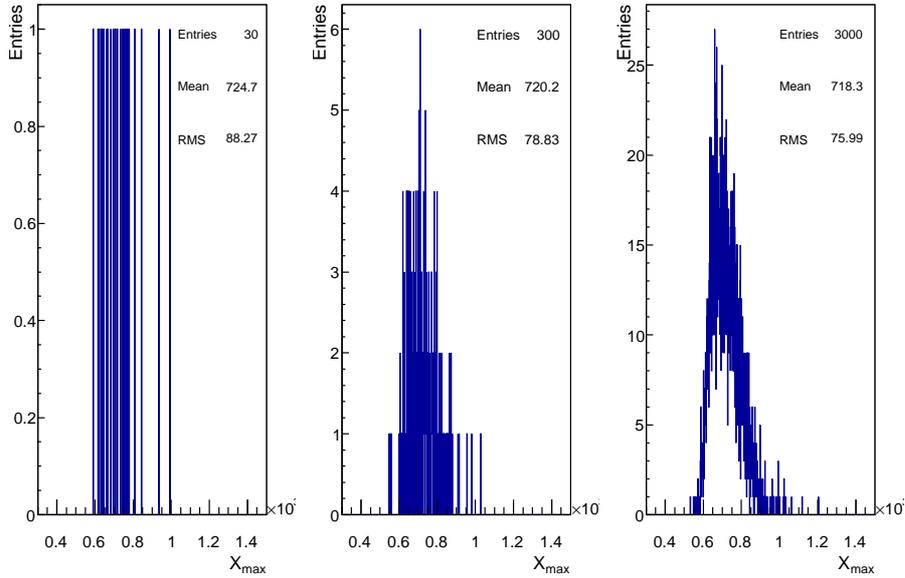}
\caption{Sample data distributions analyzed for 30, 300, and 3000 events
  corresponding to the distributions in Fig.\ref{xmaxpdfs}.}
\label{ddatas}
\end{figure}

\begin{figure}[H]
\centering
\includegraphics[scale=0.6]{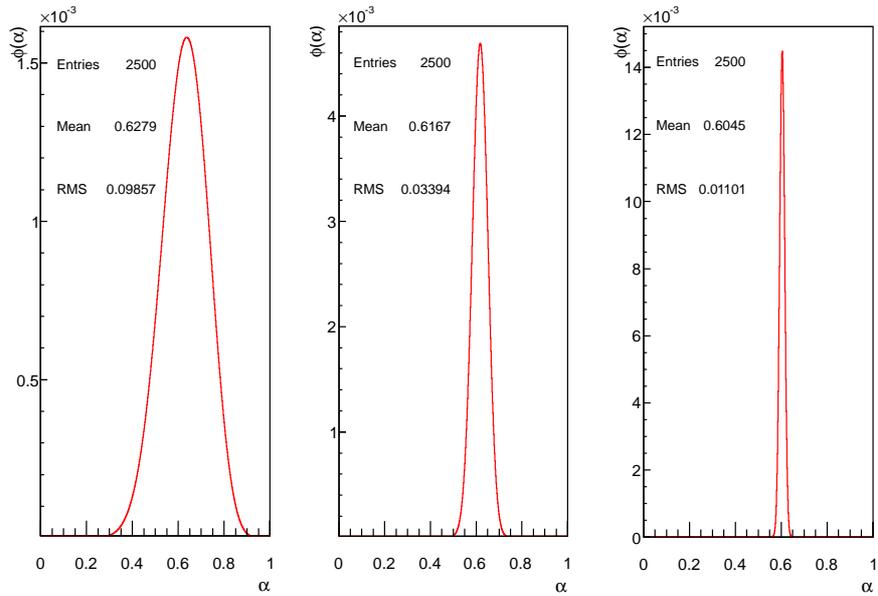}
\caption{Posterior probability functions of $\alpha$ for 30, 300, and 3000
events.}
\label{dhfs}
\end{figure}

In figure \ref{ddatas} we  show the data distributions for three sample cases,
with $\alpha = 0.6$  and in figure \ref{dhfs} we show the corresponding
posterior probability for these examples.
In table \ref{table:xmax}, we show our results for these analysis, where
the uncertainties are calculated in the following way: for the uncertainty in
the mean value we take the RMS of the posterior distribution, for the
uncertainty in the maximum likelihood value we take the width at $68\%$ 
confidence level.
\begin{table}[h]
\centering
\begin{tabular}{ccc}
\hline
\hline
\# Events & $\mean{\alpha }$ & $\alpha_{\rm max}$  \\
\hline
\hline
30 & $0.63 \pm 0.10$ & $0.64 \pm 0.10$    \\
\hline
300 & $0.62 \pm 0.03$ & $0.62\pm \pm 0.03$  \\
\hline
3000 & $0.604 \pm 0.010$ & $0.604\pm 0.010$    \\
\hline
\hline
\end{tabular}
\caption{Composition fraction obtained with the mean and maximum values of the
  posterior probability function of $\alpha$. In this case $\alpha_{true} =
  0.6$.}
\label{table:xmax}
\end{table}
In this example, both mean and maximum of the posterior probability give us
the same results.

It has been suggested that one can eliminate partially the hadronic model
dependence by shifting the mean values of the two distributions so that they
coincide \cite{Facal,Yushkov}. However we expect that by centering the
distributions one looses information to discriminate between the two
compositions. The distance between the two distributions shown in
fig. \ref{xmaxpdfs}, if we center the two distributions is $d'_1 = 0.62$ to be
compared to the actual distance $d_1= 1.55$. From our discussion, therefore,
one expects a loss of resolution of the order $s = \sqrt{d'_1/d_1} = 0.63$,
{\it i.e.} each event used in the discrimination with centering is worth a
factor $s$ less than if used without centering. This has to be compared to the
systematic uncertainty due to the hadronic model.

\section{Conclusions} 
\label{end}

We have studied different estimators for the evaluation of composition fraction
between two distributions.  We have shown that the best methods are the
maximum and the mean of the probability distribution.  The
$\chi^2$ method gives results comparable to the maximum likelihood estimator,
if the number of events is large, but by rebinning the results can be
misleading. Also, we obtained the remarkable results that with few events, the
mean value of the probability distribution is the best estimator.

We found a measure of distance between the two probabilities which gives us an
estimation of the discrimination power for two distributions. If the distance
$d_1$ is small, the discrimination between the two compositions will be
poor. If the distance is large it will be optimal. We have shown that as a
``rule of thumb'' the discrimination power scales as $1/\sqrt{d_1 N}$ with $N$
the number of events. Generalization of these methods to include an arbitrary
number of components is straightforward and will be discussed separately.

\section{Acknowledgments}

We thank J. Alvarez-Mu\~ niz, S. Riggi, I. Vali\~ no, and E. Zas for
discussions.  We thank Alexey Yushkov for discussions and for pointing to us
the relevance of centering the distributions. We thank Xunta de Galicia -
Conseller\'\i a de Educaci\'on (Grupos de Referencia Competitivos – Consolider
Xunta de Galicia 2006/51); Ministerio de Educaci\'on, Cultura y Deporte (FPA
2010-18410 and Consolider CPAN - Ingenio 2010); Ministerio de Econom\'\i a y
Competitividad (FPA2012-39489), ASPERA - AugerNext (PRI-PIMASP-2011-1154) and
Feder Funds, Spain.  We thank CESGA (Centro de SuperComputaci\'on de Galicia)
for computing resources.

\section{Appendix A: Measures of distance}
\label{appendixA}

There are many measures of distance used in the literature.
A much used measure of distance for probability distributions is the
difference between the means
\begin{equation}
d_{\bar x} = \frac{\bar x_1 - \bar x_2}{\sigma}.
\end{equation}
where $\bar x_i$ is the mean value of $x$ for the two distributions and
$\sigma$ is a measure of the width of the distributions (for instance,
$\sigma^2 = \sigma_1^2+\sigma_2^2$ has been used). It is however too
restrictive. If   $\bar x_1 =  \bar x_2$ then this distance is zero,
suggesting that the two distributions can not be discriminated. 

For two square integrable functions one can define the distance
\begin{equation}
d_2 = \int dx (g_1(x)-g_2(x))^2.
\end{equation}
This is much used in Physics, but for probability distributions is not
useful, since it is not invariant against changes of variables. 

A much used distance for probability distributions is the relative entropy
distance, also known as the Kullback-Leibler \cite{Kullback-Leibler} metric
\begin{equation}
d_{KL} = \int dx g_1(x) \log(g_2(x)) - g_2(x) \log(g_1(x)).
\end{equation}
For us, however, is not the relevant measure to use. It gives a distance of
$\infty$ if there is a no-overlapping region ( $g_2(x)= 0$ and $g_1(x) \neq
0$, for instance), which is the most relevant case in our problem.

We have found that the best choice of distance is that given by the
overlapping area
\begin{equation}
d_1 = \int dx |g_1(x) - g_2(x)|.
\end{equation}
It measures somehow the amount of probability which is not ``separable''
between the two distributions. It is bounded between 0 and 2. $d_1 = 2$
implies that the two distributions do not overlap. $d_1 = 0$ means that the
two distributions are equal.

\end{document}